\documentclass[preprint,showpacs,superscriptaddress,amsmath,amssymb]{revtex4}
\usepackage{graphicx}
\usepackage{dcolumn}
\begin{document}

\preprint{RPC}

\title{The Renormalized Tensor Interaction in a Nucleus}
\author{S.J.Q. Robinson}

\affiliation{Department of Physics, University of Southern Indiana,
  Evansville, Indiana 47712} 

\author{L. Zamick}
\affiliation{Department of Physics and Astronomy, Rutgers University,
  New Brunswick, New Jersey 08903, USA }

\date{\today}

\begin{abstract}
We show several examples were the tensor interaction of the lowest order G
matrix in a nucleus is too strong.  The examples include the quadrupole
moment of $^{6}$Li, the isosplitting of the lowest 0$^{-}$ states in
$^{16}$O, the near vanishing Gamow-Teller matrix element in the weak decay
of the J=0 T=1 state of $^{14}$O to the J=1 T=0 ground state of $^{14}$N,
and the magnitude of the deformation of $^{12}$C.  It would appear that we
could get better results by decreasing the tensor interaction strength by
about a factor of two.  We then examine the simple estimates of Gerry Brown
concerning second order tensor effects.  We note that for the triplet even
channel the combination of first and second order tensor does indeed yield
an effective weaker tensor interaction and helps to get better agreement
with experiment.
\end{abstract}

\maketitle

\section{Introduction}

In order to study the effects of the tensor and spin orbit interactions in
nuclei we use a simple interaction

\centerline{V = V$_{c}$ + x V$_{s.o.}$ + y V$_{t}$}

where c $\equiv$ central, s.o. $\equiv$ spin-orbit and t $\equiv$ tensor.  For
x=1, y=1 we select V so as to be close to a realistic G matrix like Bonn A.
We can turn the spin orbit interaction off (on) by setting x=0 (x=1).
Likewise we can turn the tensor interaction off (on) by setting y=0(1).  This allows us to study behaviors as a function of x and/or of y.

This interaction~\cite{AA} is a modification and correction of a
previous interaction~\cite{BB}. This change does not affect
calculations purely in the p shell but there are some changes when
core excitations are included. To avoid confusion we call the current
interaction V(2005) and the previous one V(1991),i.e after the year of
publication. The details and reasons for the changes are given in
ref.~\cite{AA}.

Our main thesis will be that the tensor interaction given by a bare G
matrix is too strong in the isospin T=0 channel.  By simply making it
weaker we can correlate a lot of data and be rid of a lot of
anomalies.

We conclude by presenting simple arguments by Gerry Brown to justify
using a weaker tensor interaction in the valence space.  Care must be
taken in that alternate explanations could give the same result as a
weaker tensor interaction e.g.  a stronger spin-orbit interaction.

The topics we discuss are:
\begin{enumerate}
\item The quadrupole moment of the J=1$^{+}_{1}$ state of $^{6}$Li.

\item The T=1 -- T=0 energy splitting of 0$^{-}$ states in $^{16}$0.

\item The near vanishing of the Gamow-Teller Matrix Element for A=14
$^{14}$O(J=0 T=1) $\rightarrow$ $^{14}$N(J=1 T=0).

\item The effect of the tensor interaction on Single Particle Energies in
Open Shell Nuclei - $^{12}$C and $^{14}$N.
\end{enumerate}

Some of these results have been discussed in previously~\cite{lz1}-~\cite{lz5}.

\section{The Quadrupole moment of $^{6}$Li 1$^{+}$ state}

The nucleus $^{6}$Li is often described in cluster models as a deuteron plus
an alpha particle.  However, the quadrupole moment of the deuteron is
positive $Q = 0.288$ e fm$^{2}$ whereas that of $^{6}$Li is negative,
$Q = - 0.083$ e fm$^{2}$.

That the deuteron has a quadrupole moment leads to it having a J=1$^{+}$
ground state and hence the isospin must be T=0.  Without a tensor interaction
the quadrupole moment of the deuteron would be zero.

Expanding on the work in ~\cite{lz1} we calculate the value of Q in
various model spaces and for various strengths of the tensor
interaction i.e. y as seen in Table ~\ref{tab:qval}.

\begin{table}
\begin{center}
\caption{Static quadrupole moments e fm$^{2}$ for various model spaces
  and tensor interaction strengths (y) using the bare electric charges
  $e_{p}=1.0$ and $e_{n}=0.0$. All calculations are done with the full spin-orbit strength x=1.}
\label{tab:qval}
\begin{tabular}{cccc}
Model Space & y=0 & y=0.5 & y=1.0\\
0 $\hbar \omega$  & +0.1204 & -0.0853 & -0.281 \\
0,2 $\hbar \omega$ & +0.1011 & -0.095 & -0.279 \\
0,2,4 $\hbar \omega$ & +0.0773 & -0.222 & -0.485 \\
0,2,4,6 $\hbar \omega$ & +0.06893 & -0.2788 & -0.5815\\
0,2,4,6,8 $\hbar \omega$ & +0.06568 & -0.3052 & -0.6301 \\
\end{tabular}
\end{center}
\end{table}

We see first of all that we do need a tensor interaction for Q to be
negative. However for the full strength of the tensor interaction the
Q value is too negative, a situation that is not aided by the
expansion of the model space. Indeed, as the model space grows, the
value of Q becomes increasingly negative, requiring further quenching
of the tensor interaction.

 Note that the results with our current V(2005) interaction are qualitatively different than with the previous V(1991).  With V(2005) we get the quadrupole moment of the 1+ state of $^6$Li becoming more negative with increasing $\hbar \omega$ , and getting further away from experiment. With V(1991) the opposite happened. At the zero $\hbar \omega$ level the values of Q in ref~\cite{AA} for 0,0+2 and 0+2+4 $\hbar \omega$ were respectively -0.360,-0.251 and -0.0085 e fm$^2$ .

\section {The 0$^{-}_{1}$ T=1 to 0$^{-}_{1}$ T=0 Splitting in
$^{16}$O}

In $^{16}$O the lowest 0$^{-}$ T=0 state is at an excitation energy of
10.952 MeV while the lowest 0$^{-}$ T=1 state is at 12.797 MeV i.e. an
energy splitting $\Delta$E = 1.845 MeV ~\cite{lz2,lz3}.  In the next
  table we present the values of $\Delta$E in various model spaces.

\begin{table}
\begin{center}
\caption{ The Isosplitting of the $0^-$ states in
$^{16}$O for various spin-orbit and tensor interaction strengths. }
\label{tab:split}
\begin{tabular}{|c|c|c|c|c|} \hline
\multicolumn{1}{|c|}{} & \multicolumn{1}{|c|}{} & \multicolumn{1}{|c|}
{1p-1h} & \multicolumn{1}{|c|}{1p-1h+3p-3h} & \multicolumn{1}{|c|}
{1p1h+3p-3h+2p2h}\\ \hline \hline
\multicolumn{1}{|c|}{Interaction} & \multicolumn{1}{|c|}{} & \multicolumn{1}
{|c|}{$\Delta$E MeV} &
\multicolumn{1}{|c|}{$\Delta$E} & \multicolumn{1}{|c|}{$\Delta$E} \\
\hline
\multicolumn{1}{|c|}{x=0 y=0} & \multicolumn{1}{|c|}{} & \multicolumn{1}{|c|}
{0.011} &
\multicolumn{1}{|c|}{-0.035} & \multicolumn{1}{|c|}{0.019} \\ \hline
\multicolumn{1}{|c|}{x=0 y=1} & \multicolumn{1}{|c|}{} & \multicolumn{1}{|c|}
{3.144} & \multicolumn{1}{|c|}{3.044} & \multicolumn{1}{|c|}{2.238} \\
\hline
\multicolumn{1}{|c|}{x=1 y=0} & \multicolumn{1}{|c|}{} & \multicolumn{1}{|c|}
{0.046} & \multicolumn{1}{|c|}{-0.036} & \multicolumn{1}{|c|}{0.156} \\
\hline
\multicolumn{1}{|c|}{x=1 y=1} & \multicolumn{1}{|c|}{} & \multicolumn{1}{|c|}
{2.932} & \multicolumn{1}{|c|}{2.900} & \multicolumn{1}{|c|}{1.919} \\
\hline
\multicolumn{1}{|c|}{EXPT.} & \multicolumn{1}{|c|}{1.845 MeV} & \multicolumn{1}
{|c|}{} & \multicolumn{1}{|c|}{} & \multicolumn{1}{|c|}{} \\ \hline

\end{tabular}
\end{center}
\end{table}

We see that in the absence of a tensor interaction the splitting is
negligibly small as discussed in the work of B. R.Barrett~\cite{lz6}.  In the
smallest space (1p-1h) the splitting is too large
with the full interaction 2.932 MeV.  Adding 3p-3h makes very little
difference (2.900 MeV).  Only when 2p-2h is added to 1p-1h + 3p-3h do we
get reasonably close to 1.919 MeV calculated vs. 1.845 MeV.  This is a case
where the spin-orbit interaction plays a relatively minor role.

Again, if we insist on working in the 1p-1h space we need a weaker tensor
interaction to explain the data.  We can also give results for the Bonn
interactions.  In a relativistic formulation of matter in 1p-1h Bonn A
with an effective mass m$^{*}$=930 gives $\Delta$E=3.08 MeV in the 1p-1h
space but yields 1.87 MeV in the full (1+3) h$\omega$ space (1p-1h +
3p-3h + 2p-2h).

\section {The Near Vanishing of the Gamow-Teller Matrix Element
$^{14}$O (J=0, T=1) $\rightarrow$ $^{14}$N (J=1, T=0)}

The transition $^{14}$O(J=0, T=1) $\rightarrow$ $^{14}$N(J=1, T=0)
should be an allowed Gamow-Teller decay.  But the matrix element for
this A(GT) is close to zero.  It was shown by Inglis ~\cite{lz7} that
this was not possible for a 2 hole configuration p$^{-2}$ if only a
central and spin-orbit interaction were present.  Jancovici and Talmi
~\cite{lz8} showed early-on that this near vanishing could be
explained by the presence of a two-body tensor interaction. The decay
of $^{14}$O as well as its mirror $^{14}$C was extensively studied
experimentally by Sherr et.al.~\cite{sherr}

We give the values of A(GT) for x=1 as a function of y in ~\ref{tab:GT14}

\begin{table}
\begin{center}
\caption{The Gamow-Teller matrix elements for the A=14 decay as a function of tensor strength y}
\label{tab:GT14}
\begin{tabular}{cc}
y & A(GT) \\
0 & -1.19 \\
0.25 & -0.95 \\
0.49 & 0.00 \\
1.00 & 1.38
\end{tabular}
\end{center}
\end{table}

We see that we get a vanishing of A(GT) for y = 0.49, about half of
the tensor strength needed to fit the bare G matrix like Bonn A.  This
fits in with what occurs for other cases like $J^{\pi}=0^{-}$
isosplitting in $^{16}$O.

\begin{table}
\begin{center}
\caption{Wavefunction of the 1+ state in LS coupling as a function of
  tensor strength}
\label{tab:later}
\begin{tabular}{|c|c|c|c|} \hline
\multicolumn{1}{|c|}{y} & \multicolumn{1}{|c|}{C$^{s}_{f}$} &
\multicolumn{1}{|c|}{C$^{p}_{f}$} & \multicolumn{1}{|c|}{C$^{d}_{f}$}
\\ \hline \hline
\multicolumn{1}{|c|}{0} & \multicolumn{1}{|c|}{-0.47} & \multicolumn{1}{|c|}
{0.30} & \multicolumn{1}{|c|}{0.83} \\ \hline
\multicolumn{1}{|c|}{0.25} & \multicolumn{1}{|c|}{-0.36} & \multicolumn{1}{|c|}
{0.30} & \multicolumn{1}{|c|}{0.88} \\ \hline
\multicolumn{1}{|c|}{0.49} & \multicolumn{1}{|c|}{0.09} & \multicolumn{1}{|c|}
{0.22} & \multicolumn{1}{|c|}{0.97} \\
\hline
\multicolumn{1}{|c|}{1} & \multicolumn{1}{|c|}{0.68} & \multicolumn{1}{|c|}
{0.03} & \multicolumn{1}{|c|}{0.74} \\
\hline
\end{tabular}
\end{center}
\end{table}

The 1$^{+}$ wave function changes drastically as we increase y as can be seen in Table~\ref{tab:later}. We see that as we go from y=0 to y=0.49 the $^{3}$S amplitude drops noticeably
and the $^{3}$D amplitude increases.  Of course a pure $^{3}$D state cannot
decay to $^{3}$S or $^{1}$P, which are the only two components of the J=0
ground state of $^{14}$C in the p$^{-2}$ model space.

In fairness, there is another way to get a vanishing GT.  For y=1, we
can make the spin-orbit interaction stronger.  For y=1 x=1.44 we get a
vanishing of A(GT).  Hence if we limit ourselves to this example alone
we cannot conclude that we need a weaker tensor interaction to fit the
data.  However, in other cases e.g. $^{6}$Li quadruple moment and the
0$^{-}$ splitting in $^{16}$O, one needs an effective weaker tensor interaction and simply strengthening the spin-orbit does not resolve the issues in those
nuclei.

In using Kuo's realistic matrix elements but with a 30 percent
increase in the one body spin orbit splitting, Zamick managed to get a
vanishing B(GT)~\cite{zam1}. Recently a Stony Brook - Idaho
collaboration focused on the spin-orbit interaction for the near
vanishing of B(GT) for A=14~\cite{hol1} using arguments of Brown-Rho
scaling~\cite{bro1}.

Since in the A=14 problem we get entangled in two different physics
mechanisms, namely the weakening of the tensor interaction and the
strengthening of the spin-orbit interaction, it is vital to look at
other examples where only one of the two effects is important, as an
example the 0$^-$ isospin splitting where only the tensor interaction
is important.

There are theories in which the spin-orbit interaction becomes larger e.g.
in Dirac Phenomenology with Dirac Effective Mass Ratio $m^{*}/m$ less than
one.  The spin-orbit splitting is proportional to $m/m^{*}$.

\section{The Effect of the Tensor Interaction on Single Particle
Energies in Open Shell Nuclei - $^{12}$C}

We note that for a closed major shell like $^{4}$He or $^{16}$O the
tensor interaction does not contribute to the single particle
splitting in lowest order. When ground state correlations are
allowed as in ref~\cite{AA} there are some contributions but they are small.  However, for an open shell nucleus like $^{12}$C in lowest
order we will see otherwise.

Consider the splitting $\epsilon_{p_{1/2}} - \epsilon_{p_{3/2}}$ for various
strengths of the spin-orbit interaction (x) and the tensor interaction (y).

\begin{tabbing}
xxxxxxxxxxx \= xxxxxxxxxx \= xxxxxxxxxxxxxxx \kill \\
$^{4}$He \> xy \> $\epsilon_{p_{1/2}}-\epsilon_{p_{3/2}}$ \\
\> 00 \> 0.000 \\
\> 10 \> 3.38 \\
\> 01 \> 0 \\
\> 11 \> 3.38 \\
\\
$^{12}$C \> 00 \> -0.66 \\
\> 10 \> 3.84 \\
\> 01 \> -4.75 \\
\> 11 \> -0.25 \\
\\
$^{16}$O \> 00 \> 0 \\
\> 10 \> 5.00 \\
\> 01 \> 0 \\
\> 11 \> 5.00 \\
\end{tabbing}

 Note that for $^{4}$He and $^{16}$O we get positive
splitting, 3.38 MeV and 5.00 MeV respectively but for $^{12}$C the
value is negative - 0.25 MeV.  As seen in the table the tensor
interaction gives no contribution to the splitting in $^{4}$He or
$^{16}$O but it gives a large \underline{negative} contribution for
$^{12}$C.  In this nucleus the spin-orbit and tensor interactions act
in opposite ways. That the tensor interaction acts like an opposite
sign spin-orbit force was suggested first by Wong~\cite{wong} and
Scheerbaum.~\cite{scheer}

How does all this manifest itself?  One can look at the excitation
energies of 1$^{+}$ T=0 and 1$^{+}$ T=1 states in $^{12}$C.  We will
show that energy considerations alone are misleading.  In a 1p-1h
calculation relative to a closed $p_{3/2}$ shell it is not surprising
that the excitation energies of these states is proportional to in the
spin-orbit strength.  Note however that for x=0 the two states above
are at negative energies (i.e. below the ground state) -4 MeV for the
T=0 state and -l MeV for the T=1 state.  This is due to the tensor
interaction which effectively acts as a ``negative'' spin-orbit
interaction.  But in full p shell calculations the energies of these
states is relatively flat as a function of the spin-orbit strength --
not at all linear -- from x=0 to about x=2.

This means that energy considerations are of no value.  However the transition
rates from J=0 to J=1$^{+}$ are dependent on the spin-orbit and tensor
interactions.  The J=1$^{+}$ T=0 rate is severely altered by weak isospin
admixtures.

\begin{table}
\begin{center}
\caption{ The M1 matrix element  for the excitation of a J=1+
   T=1 state in  $^{12}$C for various x and y's}
\label{tab:later2}
\begin{tabular}{|c|c|c|c|c|c|} \hline
\multicolumn{1}{|c|}{y=1} & \multicolumn{1}{|c|}{x} &
\multicolumn{1}{|c|}{A(M1)} & \multicolumn{1}{|c|}{x=1} &
\multicolumn{1}{|c|}{y} & \multicolumn{1}{|c|}{A(M1)}
\\ \hline
\multicolumn{1}{|c|}{} & \multicolumn{1}{|c|}{0.0} & \multicolumn{1}{|c|}
{0.03} & \multicolumn{1}{|c|}{} &
\multicolumn{1}{|c|}{0} & \multicolumn{1}{|c|}{1.24} \\ \hline
\multicolumn{1}{|c|}{} & \multicolumn{1}{|c|}{0.5} & \multicolumn{1}{|c|}
{0.57} & \multicolumn{1}{|c|}{} &
\multicolumn{1}{|c|}{0.5} &
\multicolumn{1}{|c|}{1.06} \\ \hline
\multicolumn{1}{|c|}{} & \multicolumn{1}{|c|}{1.0} & \multicolumn{1}{|c|}
{0.94} & \multicolumn{1}{|c|}{} & \multicolumn{1}{|c|}{1.0} &
\multicolumn{1}{|c|}{0.94} \\ \hline
\multicolumn{1}{|c|}{} & \multicolumn{1}{|c|}{1.5} & \multicolumn{1}{|c|}
{1.37} & \multicolumn{1}{|c|}{} & \multicolumn{1}{|c|}{1.5} &
\multicolumn{1}{|c|}{0.81} \\
\hline
\multicolumn{1}{|c|}{} & \multicolumn{1}{|c|}{2.0} & \multicolumn{1}{|c|}
{1.86} & \multicolumn{1}{|c|}{} & \multicolumn{1}{|c|}{2.0} &
\multicolumn{1}{|c|}{0.79} \\
\hline
\end{tabular}
\end{center}
\end{table}

Note that for x=0, y=0 we have the SU(4) limit for which the spin part of the
M1 amplitude vanishes.  We see that indeed A(M1) increases as x increases.
However A(M1) decreases as y increases supporting the fact that the tensor
interaction acts like an ``anti-spin orbit force''.

In the last few years increasing interest has developed in the topic
of the effects of the tensor interaction in open shell nuclei by
Otsuka et. al.~\cite{lz9, lz10,lz11}.  In~\cite{lz11} they also note
that a weaker tensor interaction is needed in the T=0 channel, but not
T=1.  See also the work of Brink and Stancu~\cite{lz12}.

\section{Second order Tensor contributions}

Following arguments of Gerry Brown~\cite{brown2} we make
simple evaluations of second order tensor interaction contributions to
the effective central and tensor interactions in a nucleus.

We use the Hamada-Johnson interaction~\cite{brown1} which is here given:

\begin{equation}
V_C = 0.08 \frac{1}{3} \mu (\tau _1 \cdot \tau _2) (\sigma _1 \cdot \sigma _2)
Y(x) (1 + a_C Y(x) + b_C Y^2 (x))
\end{equation}

\begin{equation}
V_T = 0.08 \frac{1}{3} \mu (\tau _1 \cdot \tau _2) Z(x) [1 + a_T Y(x) +
b_T Y^2(x)]
\end{equation}

Here the unitless parameter x is $x= \mu r$ with $\mu^{-1}=1.415$fm
and in terms of energy $\mu$=139.4 MeV

Where 
\begin{equation}
Y(x) = \frac{e^{-x}}{x}
\end{equation}
and 
\begin{equation}
Z(x)= (1 + \frac{3}{x} + \frac{3}{x^2}) Y(x)
\end{equation}

As we are interested only in the S=1 T=0 case, $\tau \cdot \tau$ will
be -3 and $\sigma \cdot \sigma$ will be 1. The coefficients are given
in Table \ref{tab:parm}.

\begin{table}
\begin{center}
\caption{Parameters} 
\label{tab:parm}
\begin{tabular}{cccccc}
S & L   & a$_C$    & b$_C$  & a$_T$ & b$_T$ \\
1 & odd & -9.07 & 3.48 & -1.29 & 0.55 \\
1 & even& 6.0 & -1.0 & -0.5 & 0.2 \\
\end{tabular}
\end{center}
\end{table}

The second order tensor effects as given by Gerry Brown are 

\begin{equation}
\delta (V_C)_{eff} = - \frac{8 V^2(r)}{e_{eff}}
\end{equation}
and 
\begin{equation}
\delta (V_T)_{eff} = \frac{2 S_{12} V^2(r)}{e_{eff}}
\end{equation}

where $e_{eff}$ is an Energy parameter ranging from 222 to 264 MeV
depending on the matrix elements under consideration. Here we choose
$e_{eff}=250$ MeV for simplicity.

The second order contribution to the central interaction is negative
definite and for the tensor it is positive definite. Hence whether
there is destructive or constructive interference depends on the sign
of the first order term as given above.

  We start from a Moszkowski-Scott cutoff radius of 1 fermi
  corresponding to x of 0.7. They argued that the part of the
  attraction up to the cut-off radius of 1 fermi cancels out the
  short-range repulsion whose range is about 0.4 fermi.~\cite{ms1}

The Moszkowski-Scott method plays an important role in the
justification of the Vlow k method of the Stony Brook
Group.~\cite{hol2}.

\begin{figure*}
\includegraphics[scale=1.5]{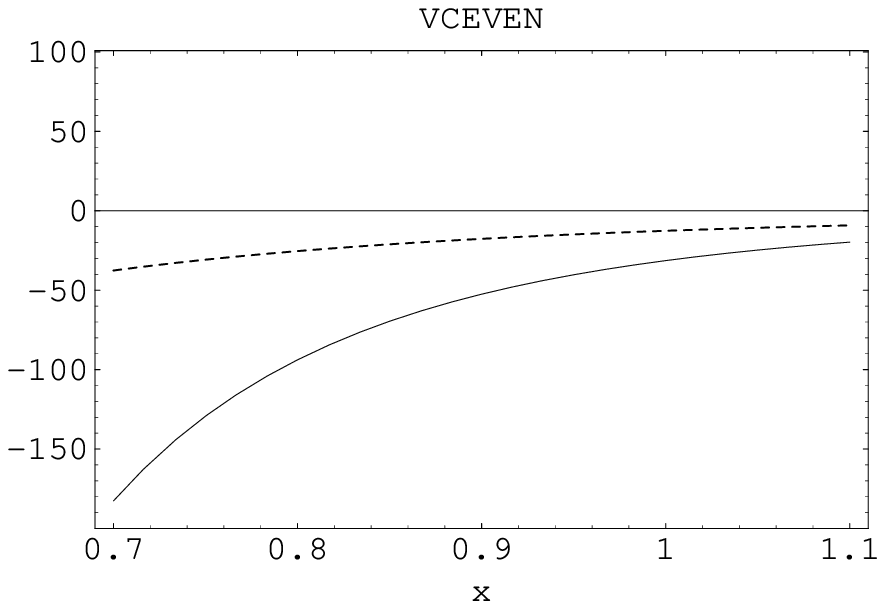}
\caption{The central force for even values of L - the dashed line is
  without the second order correction, the solid line is with that
  included.
\label{fig:vce}}
\end{figure*}

\begin{figure*}
\includegraphics[scale=1.5]{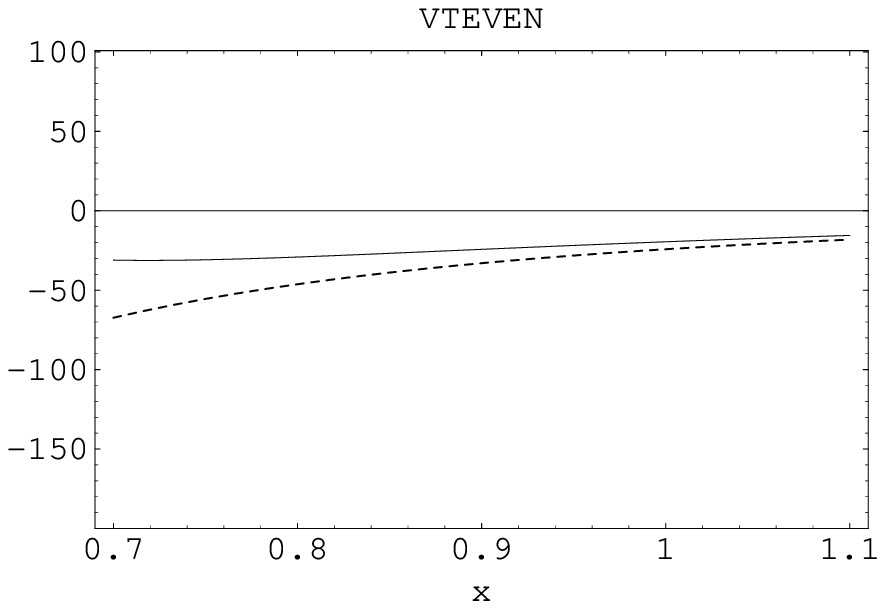}
\caption{The tensor force for even values of L - the dashed line is
  without the second order correction, the solid line is with that
  included.
\label{fig:vte}}
\end{figure*}

\begin{figure*}
\includegraphics[scale=1.5]{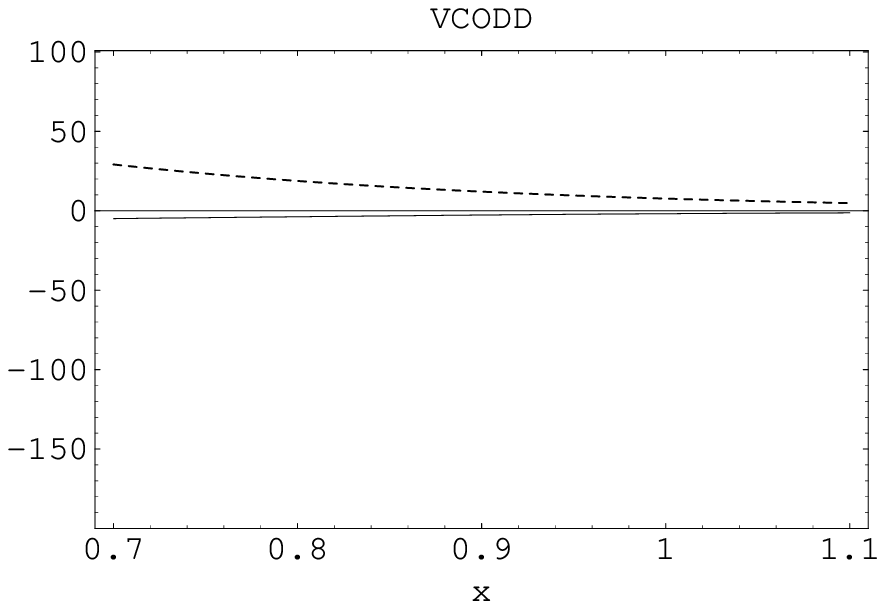}
\caption{The central force for odd values of L - the dashed line is
  without the second order corrections, the solid line is with it
  included.
\label{fig:vco}}
\end{figure*}

\begin{figure*}
\includegraphics[scale=1.5]{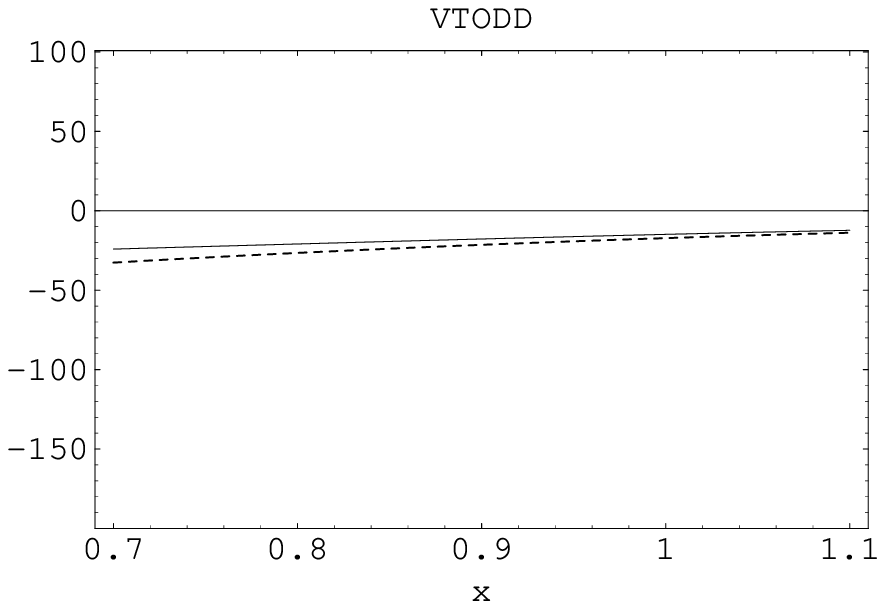}
\caption{The tensor force for odd values of L - the dashed line is
  without the second order correction, the solid line is with that
  included.
\label{fig:vto}}
\end{figure*}

In figures \ref{fig:vce} to \ref{fig:vto} we look at the cases of the
$V_{central}$ and $V_{tensor}$ for both even and odd values of L. 

The first order central even is negative so adding the second order
tensor contributions makes it more negative. This is reasonable from
the existence of the T=0 even channel bound state for two nucleons,
namely the deuteron. The bare central interaction in this channel is
not deep enough to support a bound state so the tensor interaction has
to contribute. We can see this in Figure \ref{fig:vce}

The first order tensor even interaction is negative (when the -3
factor is included) so the combination of first and second order terms
must be less negative or weaker. The sign of the first order tensor
interaction is determined phenomenologically by the positive sign of
the quadrupole moment of the deuteron. Also the sign is consistent
with the one pion exchange potential. This supports all the conclusions
of the previous sections where we see repeatedly that the bare G
matrix tensor interaction is too strong in the T=0 channel and needs
weakening.

For the odd channels we have first in Figure \ref{fig:vco} the central
odd potential. Here the attractive contribution of the second order
term pulls an initially repulsive first order term down sufficiently
so that it is slightly attractive. For the tensor odd interaction in
\ref{fig:vto}, the inclusion of the second order term again makes the
total tensor portion less attractive.

The topic of the weakening of the tensor interaction is very relevant
to the problem of nuclear pairing in the T=0 channel~\cite{moya}
especially in regarding the size of the pairing gap.

We emphasize that the main point of this work is to show that
there are clear experimental signatures that require that in the spin
triplet channel renormalization relative to a bare G matrix are
required. This is especially true for even L states where not only is
the effective central part of the interaction made deeper but also the
effective tensor interaction is weakened (screening effects). Relative
to the use of only a first order tensor interaction, the combined
first and second order tensor interaction helps to explain the the
smaller energy splitting of T=1 and T=0 0$^-$ states in $^{16}$O and the
vanishing of the Gamow-Teller matrix element in the $^{14}$C beta
decay. Also the anti-spin orbit effects in open shell nuclei like
$^{12}$C are reduced although they are still substantial.  We still
have a problem with $^{6}$Li. Although we have shown that one needs
the tensor interaction to get a negative quadrupole moment we get it
to be increasingly negative with increasing model space.

In closing we note that the shell model works very well in the p shell
as noted by the many works of Cohen and Kurath~\cite{kur1}. One
purpose here is to understand why it works so well. We see that
although the explicit configurations involving higher shells are not
present in most calculations, their implicit presence is of crucial
importance for the success of the model. We have adopted a low-tech
approach to illustrate this point. For more trustworthy quantitative
results, the high powered ab initio shell model methods or other
equivalent methods need to be employed.  Nevertheless we feel that the
more qualitative methods used here are of considerable value in providing
insight into the physics behind these more complex approaches.

\begin{thebibliography}{00}

\bibitem{AA}L.Zamick M.S. Fayache and Y.Y. Sharon
Annals of Physics, Volume 317, Issue 1, May 2005, Pages 148-151

\bibitem{BB} D. C. Zheng and L. Zamick Annals of Physics, Volume 206,
  Issue 1, 15 February 1991, Pages 106-121

\bibitem{lz1} L. Zamick, D. C. Zheng and M. Fayache PRC
  \underline{51}, (1995) 1253. 

\bibitem{lz2} D. Zheng, L. Zamick, M. S. Fayache and H. Muther, Annals
  of Physics \underline{230}, (1994) 118. 

\bibitem{lz3} D. C. Zheng and L. Zamick, Annals of Physics,
  \underline{206}, (1991) 106. 

\bibitem{lz4} D. C. Zheng, L. Zamick and H. Muther, Ann. Phys. NY
  \underline{230} 118 (1994) 

\bibitem{lz5} The Nuclear Tensor Interaction, M. S. Fayache, L. Zamick
  and B. Castel, Physics Reports 290 (1997) 201-282. 

\bibitem{lz6} B. R. Barrett, thesis and Phys Rev 159 (1967) 816. 

\bibitem{lz7} D. R. Inglis, Rev. Mod. Phys. \underline{25}, (1953) 390. 

\bibitem{lz8} B. Jancovici and I. Talmi, Phys. Rev. \underline{95},
  (1954) 289. 

\bibitem{sherr} R. Sherr,J.B.Gebhart,H.Horie and W.F. Hornyak,
  Phys.Rev.100 ,Letter to the editor,945 (1955)

\bibitem{zam1} L. Zamick, Phys Lett 21 (1966) 194.

\bibitem{hol1} J.W. Holt, G.E. Brown, T.T.S. Kuo, J.D. Holt, and
  R. Machleidt, Lanl arXiv nucl-th 07010.0310

\bibitem{bro1} G.E. Brown and M. Rho, Phys. Rept. 396 (2004) 1.

\bibitem{wong} C.W. Wong Nucl. Phys. A 108 (1968) 481.

\bibitem{scheer} R.B. Scheerbaum, Phys. Lett. B63 (1976) 381.

\bibitem{lz9} T. Otsuka, R. Fujimoto, Y. Utsuno, B. A. Brown, M. Hanna
  and T. Mizusaki, Phys. Rev. Lett. \underline{87}, 082502 (2001). 

\bibitem{lz10} T. Otsuka, T. Suzuki, R. Fujimoto, H. Grawe,
  Y. Akaishi, Phys. Rev.  Lett. \underline{95}, 232502 (2005). 

\bibitem{lz11} B. A. Brown, T. Duguet, T. Otsuka, D. Abe and
  T. Suzuki, Phys. Rev.  C\underline{74}, 061303(R) (2006). 

\bibitem{lz12} D. M. Brink and F. L. Stancu,
  Phys. Rev. C\underline{75}, 064311 (2007).


\bibitem{brown2} G. E. Brown, Unified Theory of Nuclear Models and
  Forces, North-Holland Publishing Company, Amsterdam (1971).

\bibitem{brown1} G. E. Brown and A. D. Jackson, \underline{The
  Nucleon-Nucleon Interaction}, North-Holland Publishing Company,
  Amsterdam, (1976). 

\bibitem{ms1} S.A. Moszkowski and B. Scott Ann. of Physics, 11,(1960) 65

\bibitem{hol2} J.W. Holt and G.E. Brown lanl arXiv nucl-th 10408047V1.

\bibitem{moya} E. Garrido, P. Sarriguren, E. Moya de Guerra,
  U. Lombardo, P. Schuk, and H.J. Schultze Phys. Rev. C 63, 037304.

\bibitem{kur1} S. Cohen and D. Kurath Nucl. Phys. 73, 1 (1965).

\end{thebibliography}
\end{document}